\documentclass{aa}
\usepackage{graphics}
\usepackage{latexsym}
\begin{document}

   \thesaurus{22     
              (13.07.2;  
               03.20.9;  
               03.20.8;  
               03.20.1;  
               03.09.1)}  
   \title{Event reconstruction in high resolution Compton telescopes}

   \author{S. E. Boggs\inst{1}
          \and
          P. Jean\inst{2}}

   \offprints{S. E. Boggs}

   \institute{Space Radiation Laboratory, California Institute of Technology,
              MC 220-47, Pasadena CA, 91125 USA\\
              email: boggs@srl.caltech.edu
         \and
             Centre d'Etude Spatiale des Rayonnements,
             UPS-CNRS, Toulouse, France\\
             email: Pierre.Jean@cesr.fr
             }

   \date{Received April 6, 2000; accepted May 9, 2000}

   \titlerunning{Compton Event Reconstruction}
   \authorrunning{Boggs et al.}
   \maketitle

   \begin{abstract}

The development of germanium Compton telescopes for nuclear
$\gamma$-ray astrophysics ($\sim$0.2-20~MeV) requires
new event reconstruction techniques to accurately determine the
initial direction and energy of photon events,
as well as to consistently reject background events.
This paper describes techniques for event reconstruction, accounting
for realistic instrument/detector performance and uncertainties.
An especially important technique is Compton Kinematic Discrimination,
which allows proper interaction ordering and background rejection with
high probabilities. The use of these techniques are crucial for the
realistic evaluation of the performance and sensitivity of any
germanium Compton telescope configuration.
   
   \end{abstract}

%

\section{Compton telescopes for $\gamma$-ray astrophysics}

Looking beyond the INTErnational Gamma-Ray Astrophysics Laboratory 
(INTEGRAL), the next generation soft $\gamma$-ray ($\sim$0.2-20~MeV) observatory 
will require high angular and spectral resolution imaging to significantly 
improve sensitivity to astrophysical sources of nuclear line emission.
Building upon the success of COMPTEL/CGRO (\cite{schon93}), and the high
spectral resolution of the upcoming SPI/INTEGRAL (\cite{vedre98,licht96}), a number 
of researchers (\cite{johns96, jean96, boggs98}) have discussed the
merits of a high spectral/angular resolution
germanium Compton telescope (GCT); the ability to 
achieve high sensitivity to point sources while maintaining a large
field-of-view make a high resolution Compton telescope an attractive
option for the next soft $\gamma$-ray observatory.

The development of Compton telescopes began in the 1970's, with work 
done at the Max Planck Institut (\cite{schon73}), 
University of California, Riverside (\cite{herzo75}), and the University of 
New Hampshire (\cite{lockw79}), culminating in the design and flight of 
COMPTEL/CGRO. These historical Compton telescopes consist of two
scintillation detector planes -- a low atomic number `converter' and a high atomic 
number `absorber.' The model interaction of a Compton telescope is a single 
Compton scatter in the converter plane, followed by photoelectric absorption of 
the scattered photon in the absorber. By measuring the position and energy of 
the interactions, the event can be \textit{reconstructed} to determine the initial photon 
direction to within an annulus on the sky.

A handful of groups are actively developing imaging germanium
detectors (GeDs) partly in anticipation of a GCT (\cite{luke94, kroeg96}).
The goal of these 
researchers is to develop large area detectors with (sub)millimeter spatial
resolution, while maintaining the high spectral resolution
($E/\delta E \sim 500$ at 1~MeV) 
characteristic of GeDs. The use of high spectral/spatial resolution GeDs as
converter and absorber planes would significantly improve the performance of a 
Compton telescope, but will add a number of complications to the event
reconstruction. Most significantly, with the moderate atomic number $(Z = 32)$ of
germanium, photons will predominantly undergo multiple Compton scatters 
before being photoabsorbed in the instrument. Furthermore, with interaction 
timing capabilities of $\sim$10~ns, the interaction order will not be determined 
unambiguously by timing alone. Compton Kinematic Discrimination (CKD) is 
proposed here to overcome these complications, an extension of a method first 
discussed in context of liquid xenon time projection chambers (\cite{april93}).
The ability of this technique to allow proper event reconstruction is
investigated in detail.

Due to their relatively low efficiency (typically $\sim 1\%$), 
Compton telescopes rely on efficient background suppression to maintain their 
sensitivity. In addition to interaction ordering, techniques are presented using 
CKD, in combination with other tests and restrictions,
to suppress the dominant background components.

The goal of this work is to outline a complete set of event
reconstruction techniques for GCTs, taking into account realistic
detector/instrument performance and uncertainties. Examples of
the techniques are presented for a GCT configuration outlined in
Appendix~A; however, full analysis of this configuration
will be presented in a second paper dedicated to the
optimization and performance of several GCT configurations.
The full analysis of a GCT configuration is complicated,
requiring a detailed study of the tradeoffs between efficiency,
angular and spectral resolution; therefore, this paper focuses
only on the detailed discussion of the event reconstruction
techniques which will be used in future work dedicated to
analyzing GCT performance.

\begin{figure}
  \resizebox{\hsize}{!}{\includegraphics{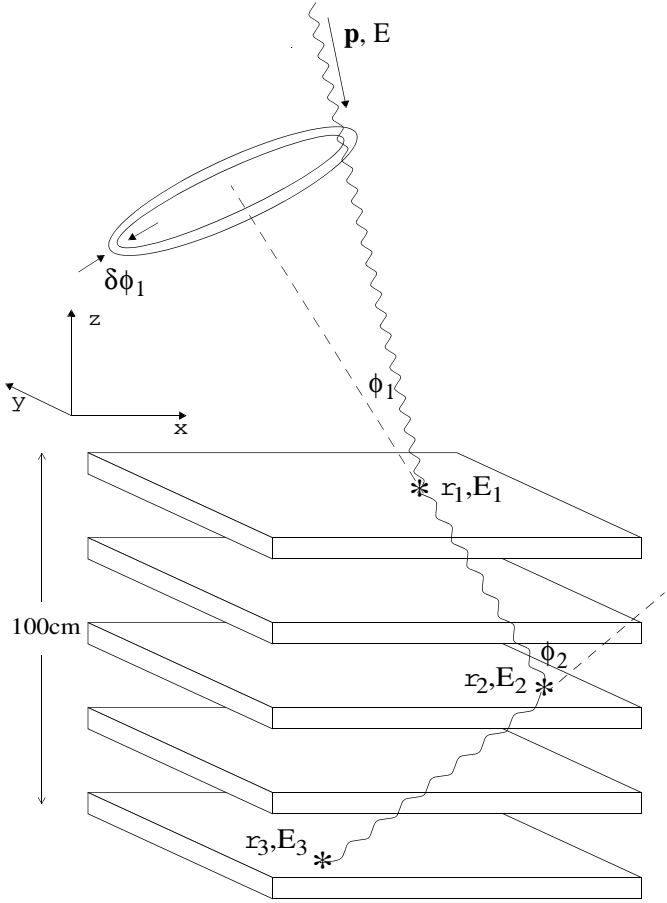}}
  \caption{Example Compton telescope. If a photon undergoes one or more
   Compton scatters in the instrument and then is photoelectrically absorbed,
   then by using the positions $(\vec{r}_{1}, \ldots,\vec{r}_{N})$ and
   energy deposits $(E_{1}, \ldots, E_{N})$ of the interactions, the
   initial direction of the photon can be determined by the Compton scatter
   formula to within an annulus on the sky, $\phi_{1}$. The width of this
   annulus is determined by the uncertainties in both the interaction
   locations and energy deposits.}
  \label{Figure 1}
\end{figure}
\section{Principles of Compton imaging}

The principle of Compton imaging of $\gamma$-ray photons is illustrated in 
Figure~1. (See von Ballmoos, Diehl and Sch\"onfelder (1989)
for an excellent review of historical Compton telescope configurations.)
An incoming photon of energy $E$ and direction $\vec{\hat{p}}$ undergoes a Compton
scatter at an angle $\phi_{1}$ at the position $\vec{r}_{1}$ within a detector,
creating a recoil electron of energy of $E_{1}$ which is quickly absorbed
and measured by the detector itself. The scattered photon then deposits the
rest of its energy in the instrument in a series of one or more interactions
of energies $E_{i}$ at the positions $\vec{r}_{i}$, 
until eventually photoabsorbed. Here the total photon energy after each scatter 
$i$, normalized to the electron mass, is defined as

\begin{equation}
W_{i}=\frac{1}{m_{e}c^{2}} \sum_{j=i+1}^{N} E_{j} ~,
\end{equation}
where $W_{0}=E/m_{e}c^{2}$, and $N$ is the total number of interactions. The initial
photon direction is related to scatter direction vector
$\vec{r}_{1}'=\vec{r}_{2}-\vec{r}_{1}$
($\vec{\hat{r}}_{1}'$ after normalization), and the scattered 
photon energies $W_{i}$ by the Compton formula

\begin{equation}
\vec{\hat{r}}_{1}' \cdot \vec{\hat{p}} = \cos{\phi_{1}} =1 + \frac{1}{W_{0}} - \frac{1}{W_{1}} ~.
\end{equation}

Given the measured scatter direction $\vec{\hat{r}}_{1}'$ and
the angle $\cos{\phi_{1}}$ implied from the 
energy depositions, the equation for $\vec{\hat{p}}$ is not unique (if the electron recoil
direction could be measured, it would solve this ambiguity); therefore, the initial 
direction of the photon cannot be determined directly, but it can be limited to an 
annulus of directions $\vec{\hat{p}}'$ which satisfy the equation

\begin{equation}
\vec{\hat{r}}_{1}' \cdot \vec{\hat{p}}' = \cos{\phi_{1}} ~.
\end{equation}

There are two uncertainties in determining the event annulus: the
uncertainty in $\phi_{1}$ due to the finite energy resolution of the detectors, here labelled 
$\delta \phi_{1,E}$, and the uncertainty in $\vec{r}_{1}'$ determined
by the spatial resolution of the 
detectors. Both of these uncertainties add to determine the uncertainty (effective width)
of the event annulus $\delta \phi_{1}$. From Equation~2, the derivation
of $\delta \phi_{1,E}$ is straightforward and yields

\begin{equation}
\delta \phi_{i,E} = \frac{1}{\sin{\phi_{i}}}
[( \frac{\delta W_{i-1}^{2}}{W_{i-1}^{4}}) +
\delta W_{i}^{2} (( \frac{1}{W_{i}^{2}} - \frac{1}{W_{i-1}^{2}})^2 -
\frac{1}{W_{i-1}^{4}})]^{1/2} ~,
\end{equation}
where,

\begin{equation}
\delta W_{i} = \frac{1}{m_{e}c^{2}} [\sum_{j=i+1}^{N} \delta E_{j}^{2}]^{1/2} ~.
\end{equation}

\begin{figure}
 \rotatebox{270}{
  \resizebox{!}{\hsize}{\includegraphics{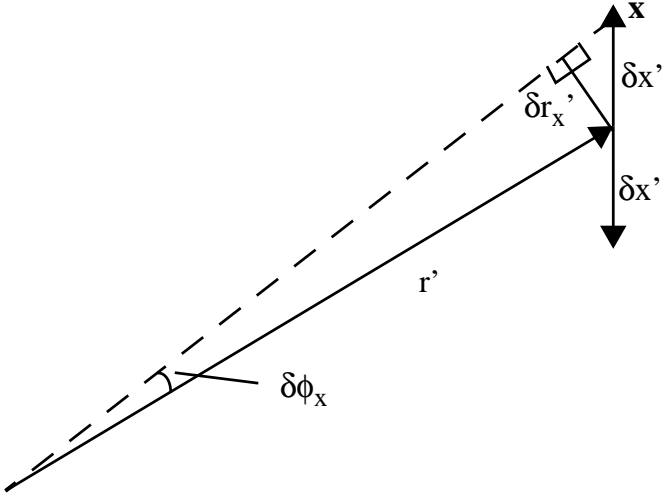}}
 }
  \caption{Illustration of how uncertainties in the scattered
   photon direction $\vec{r}'$ can be translated into 
   effective uncertainties in the scattering angle $\phi$.}
  \label{Figure 2}
\end{figure}

In order to simplify the analysis, it is convenient to transform the
uncertainty $\delta \vec{r}_{1}'$ into an effective uncertainty in $\phi_{1}$,
defined as $\delta \phi_{1,r}$, such that

\begin{equation}
\delta \phi_{1} = \sqrt{\delta \phi_{1,E}^{2} + \delta \phi_{1,r}^{2}} ~.
\end{equation}
The angular resolution $\delta \phi_{i,r}$ is the effective `wiggle'
of $\vec{\hat{r}}_{i}'$ around its measured 
direction due to the uncertainties in the spatial measurements. The spatial 
uncertainties are defined as
$\delta x'_{i} = \sqrt{\delta x_{i}^{2} + \delta x_{i+1}^{2}}$,
$\delta y'_{i} = \sqrt{\delta y_{i}^{2} + \delta y_{i+1}^{2}}$,
$\delta z'_{i} = \sqrt{\delta z_{i}^{2} + \delta z_{i+1}^{2}}$.
It is simplest to analyze the situation for each axis separately as shown in
Figure~2. The uncertainty in the direction of $\vec{\hat{r}}_{i}'$ due to the uncertainty
$\delta x'_{i}$ is given by

\begin{equation}
\delta \phi_{i,x} \simeq \tan(\delta \phi_{i,x}) = (\frac{\delta x'_{i}}{r_{i}'})\sqrt{1-(\vec{\hat{r}}_{i}' \cdot \vec{\hat{x}})^2} ~.
\end{equation}
Likewise for the other axis,

\begin{displaymath}
\delta \phi_{i,y} \simeq (\frac{\delta y'_{i}}{r_{i}'})\sqrt{1-(\vec{\hat{r}}_{i}' \cdot \vec{\hat{y}})^2} ~,
\end{displaymath}
\begin{displaymath}
\delta \phi_{i,z} \simeq (\frac{\delta z'_{i}}{r_{i}'})\sqrt{1-(\vec{\hat{r}}_{i}' \cdot \vec{\hat{z}})^2} ~,
\end{displaymath}
which combine to yield the total uncertainty $\delta \phi_{i,r}$ given by
\begin{equation}
\delta \phi_{i,r} =
\sqrt{\delta \phi_{i,x}^{2} + \delta \phi_{i,y}^{2} + \delta \phi_{i,z}^{2}} ~.
\end{equation}

For detectors with a given energy resolution,
in order to optimize the performance of a Compton telescope one would require that
$\delta \phi_{i,r} \leq \delta \phi_{i,E}$
in the energy range of interest. To first order, this implies that the spatial 
resolution in relation to the scale size of the instrument must be comparable to 
or less than the energy resolution, i.e. $\delta r_{1}'/r_{1}' \leq \delta E_{1}'/E_{1}'$.

\section{Complications of germanium Compton telescopes}

Finite detector thresholds, energy resolutions, and spatial resolutions
produce systematic biases in the imaging capabilities of Compton telescopes. 
These limitations have been discussed in detail elsewhere in context of
two-layer, low-Z converter and high-Z absorber, scintillation detector designs
(\cite{vonba89}), and the conclusions can be directly 
applied to GCTs. However, GCT designs will introduce additional
complications which significantly alter the event reconstruction techniques. Historical 
Compton telescope configurations make two assumptions about the events 
which do not generally hold in GCTs: (i) the events are a single Compton
scatter in the converter, followed by photoelectric absorption in the absorber, and 
(ii) the time-of-flight (TOF) between the photon interactions is measured to 
determine their order.

The distributions of number and type of interaction sites in a GCT for 
normally incident, fully-absorbed photons ranging from 0.2-10~MeV are 
shown in Figure~3, for the instrument configuration discussed in Appendix~A.
Here we distinguish three event types: a single photoelectric absorption,
one or more Compton scatters followed
by a single photoabsorption, and one or more pair productions.
Compton scatters followed by pair production could potentially be
reconstructed; however, here we include these events with other
pair productions. 
These distributions account for the finite spatial resolution of the detectors, so 
that interactions occurring too closely together are not resolved. From these 
distributions it is clear that events with $\sim$8 or more interaction sites can be 
immediately rejected as probable pair production events, with little effect on 
the Compton photopeak efficiency. For incident photon energies above 0.5~MeV,
3-7 interaction site Compton scatter events dominate the photopeak.

To accurately reconstruct a Compton scatter event, the first and second 
interaction sites must be spatially resolved, and their order correctly 
determined. The need to determine the proper ordering of
three or more (3+) interaction sites is
complicated by the timing capabilities of GeDs. In the scintillation detectors of 
COMPTEL/CGRO the interaction timing can be performed to $\sim$0.25~ns
(\cite{schon93}), which is adequate to determine the TOF between two
interactions in the separate detector planes. With the slower rise time of GeDs one 
can reasonably expect event timing to $\sim$10~ns, which is inadequate for TOF 
measurement in reasonably-sized instruments. While Pulse Shape
Discrimination methods have been proposed to push the interaction timing in GeDs to 
$\sim$1~ns (\cite{boggs98}), even this timing would be unreliable for determining TOF 
among three or more interaction sites. A method of reliably determining the photon
interaction order without timing information must be developed.

\begin{figure}
  \resizebox{\hsize}{!}{\includegraphics{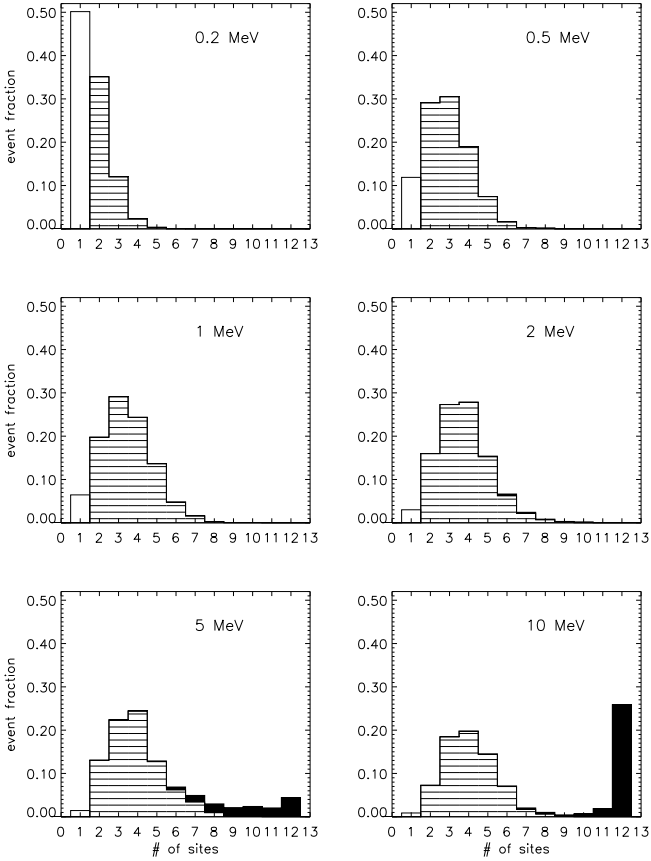}}
  \caption{Statistical distributions of the number of interaction sites for
   fully-absorbed photons for the instrument configuration presented in Appendix~A. 
   These distributions take into account the finite spatial distribution of the detectors, 
   combining interactions that cannot be spatially resolved. Events are divided 
   into photoelectric absorptions (solid white), pair productions (solid black), and 
   the desired Compton scatter(s) followed by a single photoabsorption 
   (striped). Events with 12 or more sites were combined in a single bin.}
  \label{Figure 3}
\end{figure}

\section{Multiple Compton scatter events: Compton kinematic discrimination}

One method has been suggested to overcome these complications in the 
context of liquid xenon time projection chambers (\cite{april93}). Here, 
this method is formalized as Compton Kinematic Discrimination (CKD) and 
examined in more detail. This technique allows the order of the photon
interactions to be determined with high probability, as well as providing the basis of a 
powerful tool for background suppression in GCTs.

CKD takes advantage of redundant measurement information in an event 
to determine the most likely interaction sequence. A photon of initial energy $E$ 
(using the notation in Section~2) interacts in the instrument at $N$ sites, 
depositing an energy of $E_{i}$ at each location $\vec{r}_{i}$. It is assumed that the 
interactions $1, \ldots ,N-1$ are Compton scatters, and interaction $N$ is the final 
photoabsorption. Given the correct ordering of the interactions, there are two 
independent ways of measuring $N-2$ of the
scattering angles, $\cos{\phi_{2}}, \ldots , \cos{\phi_{N-1}}$.

\textit{Geometrical measurement of }$\cos{\phi_{i}}$. From simple vector analysis, given 
the correct ordering of the interaction sites one can derive the scatter angles

\begin{equation}
\cos{\phi_{i}} = \vec{\hat{r}}_{i}' \cdot \vec{\hat{r}}_{i-1}', i = 2, \ldots,N-1 ~,
\end{equation}
where the uncertainties in the scattering angles, $\delta \cos{\phi_{i}}$, can be estimated 
from the spatial uncertainty in the scattering angles (Equation~8), 
yielding
\begin{equation}
\delta (\cos{\phi_{i}}) = \delta \phi_{i,r} \sin{\phi_{i}} ~.
\end{equation}

\textit{Compton kinematics measurement of }$\cos{\phi_{i}'}$. Given the correct ordering, 
the measured values of $W_{i}$ can be derived, which were defined earlier as the 
energy of the photon after each scattering $i$, in units of $m_{e}c^{2}$. The Compton 
scatter formula (Equation~2) gives:

\begin{equation}
\cos{\phi_{i}'} = 1+\frac{1}{W_{i-1}}-\frac{1}{W_{i}}, i = 1, \ldots,N-1 ~,
\end{equation}
\begin{equation}
\delta (\cos{\phi_{i}'}) =
[(\frac{\delta W_{i-1}^{2}}{W_{i-1}^{4}})+
\delta W_{i}^{2}((\frac{1}{W_{i}^{2}}-\frac{1}{W_{i-1}^{2}})^2-\frac{1}{W_{i-1}^{4}})]^{1/2} ~.
\end{equation}

Given the $N-2$ independent measurements of $\cos{\phi_{2}}, \ldots ,\cos{\phi_{N-1}}$,
a trial 
ordering of the interaction sites can be tested for consistency. If the assumed 
ordering is incorrect $\cos{\phi_{i}}$ will not equal $\cos{\phi_{i}'}$ in general.
Every possible 
permutation of orderings can be tested to determine the one most consistent 
with $\cos{\phi_{i}} = \cos{\phi_{i}'}$. Given a trial ordering,
the two angle cosines for sites
$i = 2, \ldots ,N-2$ are relabelled for convenience
$\eta_{i} = \cos{\phi_{i}}$, $\eta_{i}' = \cos{\phi_{i}'}$.

As a first test, trial orderings that produce values of
$\mid \eta_{i}' \mid \geq 1$ are ruled out, 
since $\cos{\phi_{i}'} < 1$ for any scattering angle $\phi_{i}'$.
This condition will eliminate 
many orderings which cannot physically be due to multiple Compton scatters 
followed by photoabsorption. Next a least-squares statistic measuring the 
agreement between the redundant scatter angle measurements is defined:

\begin{equation}
\chi^{2} = \frac{1}{N-2}
\sum_{i=2}^{N-1} \frac{(\eta_{i}-\eta_{i}')^{2}}{(\delta \eta_{i}^{2} + \delta \eta_{i}'^{2})} ~.
\end{equation}

In general, $\chi^{2}$ will be minimized when the interactions are properly ordered 
(i.e. the order in which they occurred). Therefore, all possible permutations 
can be tested for their value of $\chi^{2}$, and the ordering corresponding to the 
minimum value, $\chi^{2}_{min}$, is taken as the most likely ordering.

This consistency statistic also provides a powerful tool for rejecting 
background events. If the event is truly a multiple Compton scatter event 
followed by a photoabsorption then  $\chi^{2}_{min} \sim 1$. By setting a maximum 
acceptable level for $\chi^{2}_{min}$, events that do not fit this scenario can be 
rejected. Such events include partially-deposited photons which scatter out 
of the instrument (Compton continuum), photon interactions with spatially 
unresolved interaction sites, events with interactions below the detector 
threshold, pair-production events, and similarly $\beta^{+}$ decays. These events 
frequently have $\chi^{2}_{min} \gg 1$, allowing a strong rejection statistic that is not very 
sensitive on the level set on $\chi^{2}_{min}$. Here, $\chi^{2}_{min}$ has been treated 
as a normal least-squares statistic with $N-2$ degrees of freedom, and events are 
rejected which have probabilities of $\chi^{2}_{min} < 5\%$.
Variations in the level between $1\%$ and $10\%$ do not strongly
affect CKD rejection capabilities.
For example, varying this level from $5\%$ to $1\%$ shifted the CKD efficiency
curves in Figure~4 by 1-2\%.

\begin{figure}
  \resizebox{\hsize}{!}{\includegraphics{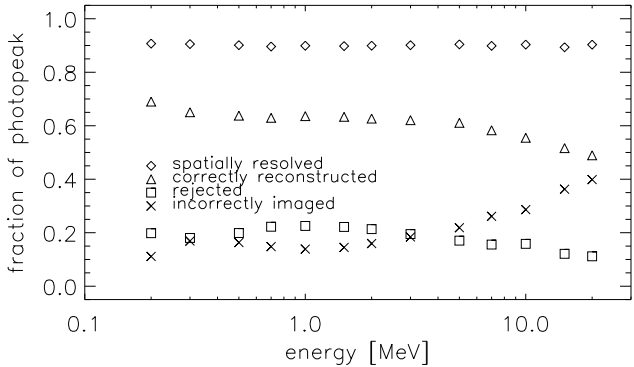}}
  \caption{Photopeak distributions for 3+ site events.
   The fraction of events with the first and second interaction
   sites spatially resolved are presented ($\diamond$), along with
   the fraction of events which have been properly ordered (hence imaged)
   using CKD ($\triangle$). The fraction of events rejected 
   by the CKD statistic ($\Box$) as well as the fraction incorrectly
   imaged ($\times$) are also shown.}
  \label{Figure 4}
\end{figure}

The fraction of 3+ site photopeak events which have the first and second 
interaction sites spatially resolved -- and hence could be imaged to the proper 
direction -- is shown in Figure~4 as a function of energy, for the instrument 
model discussed in Appendix A.
Roughly $90\%$ of all events from 0.2-20~MeV have their first and second sites
spatially resolved from each other.
(Some of these events do not have their second, third, etc.,
interactions spatially resolved from each other, and will be
rejected by the limits on $\chi^{2}_{min}$.)
This figure also shows the fraction of the photopeak events which 
CKD properly orders (correctly reconstructed),
as well as the fractions improperly ordered (hence 
incorrectly imaged to off-source background), and the fraction completely rejected. For 
energies below $\sim$10~MeV, CKD allows proper reconstruction (hence imaging) 
of $\sim 60-70\%$ of the photopeak events, while rejecting $10-20\%$. The remaining 
$10-20\%$ are incorrectly imaged into the off-source background. For 
comparison, if the order of the interaction sites were randomly chosen $<15\%$ 
would be correctly imaged, while the remaining $>85\%$ would be incorrectly 
imaged into the background.

\section{Single Compton scatter events: single scatter discrimination}

CKD will only work for $N > 2$ since there are no independent scattering 
angle measurements for a single Compton scatter followed by a 
photoabsorption. It turns out that the ordering of two-site photopeak events can 
still be determined with a high probability; however, the ability to reject 
background events is lost. As is discussed further in Section~6,
the loss of background rejection, coupled with
low peak-to-Compton ratios, and a larger fraction of backscatter events
mean that the inclusion of two-site events will likely hurt the
sensitivity of a GCT; however, discussion of event ordering is still
included for completeness.

Given a two-site event, the first test one can perform is to determine 
whether both possible orderings of the interaction sites are energetically 
compatible with a single Compton scatter, i.e. are compatible with the 
requirement that $\cos{\phi_{1}} < 1$ (Equation~2). In Figure~5, the fraction of 
spatially-resolved, photopeak events with unique orderings are plotted versus 
energy. Also plotted are the fraction with ambiguous orderings. At energies 
below $\sim$0.4~MeV the majority of resolved photopeak events have a unique 
ordering, while at higher energies most events are ambiguous.

As an empirical test of the ambiguous events, the relative magnitude of 
the energy lost in the initial scatter $(E_{1})$ compared to the
photoabsorption $(E_{2})$ can be 
compared. The fraction of resolved two-site photopeak events which have 
ambiguous orderings with $E_{1} > E_{2}$ is plotted in Figure 5. At higher energies, 
nearly all of the resolved photopeak events with ambiguous interaction orders 
have $E_{1} > E_{2}$, which can be used to determine the most likely interaction order. 
This empirical result can be easily understood, in hindsight, by the fact that 
photons which deposit most of their energy in the initial Compton scatter are 
much more likely to be photoabsorbed in the second interaction.

\begin{figure}
  \resizebox{\hsize}{!}{\includegraphics{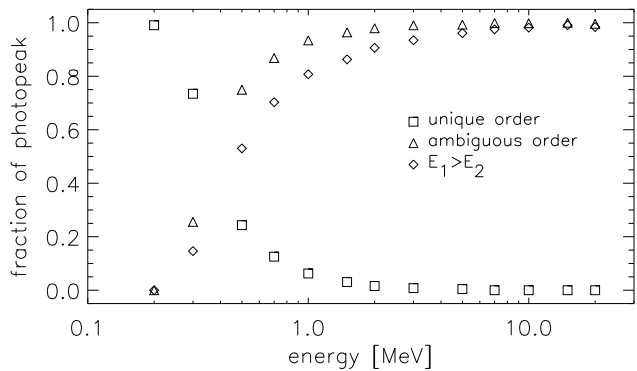}}
  \caption{Empirical distributions for fully-resolved, two-site photopeak events. 
   The fraction of events with only one physically possible ordering ($\Box$), as 
   well as events with both orderings physically possible ($\triangle$) are shown. 
   Also shown are the events with both orderings possible, but with the larger 
   energy deposit in the initial scatter rather than the photoabsorption,
   $E_{1} > E_{2}$ ($\diamond$).}
  \label{Figure 5}
\end{figure}

Therefore, a simple Single Scatter Discrimination (SSD) technique to 
determine the most likely interaction ordering of two-site events follows. First 
one determines whether a physically unique ordering exists; if not, the larger 
energy deposit is assumed to be the initial Compton scatter. Only at the lowest 
energies are two-site events possible in which neither ordering is acceptable 
(unresolved events, Compton continuum), and some background rejection is 
possible.

The fraction of two-site photopeak events which are spatially resolved is 
shown in Figure~6 as a function of energy, for the instrument model in 
Appendix~A. Roughly $80\%$ of all events from 0.2-20~MeV are resolved. This 
number is about $10\%$ lower than the 3+ site events, due to the smaller path 
lengths of the lower energy scattered photons in two-site photopeak events. 
Also shown in Figure~6 is the fraction of events which SSD has properly reconstructed.
SSD allows proper reconstruction 
(hence imaging) of $\sim 60-80\%$ of the photopeak events, while improperly 
imaging the remaining $\sim 20-40\%$ into the off-source background. Only a 
relatively small number of low energy events can be rejected outright. SSD is 
least effective around 0.5~MeV, where the unique/ambiguous ordering 
signatures are not as clear. For comparison, if the order of the interaction sites 
were randomly chosen $\sim 40\%$ would be properly imaged, while the remaining 
$\sim 60\%$ would be improperly imaged into the background.

\begin{figure}
  \resizebox{\hsize}{!}{\includegraphics{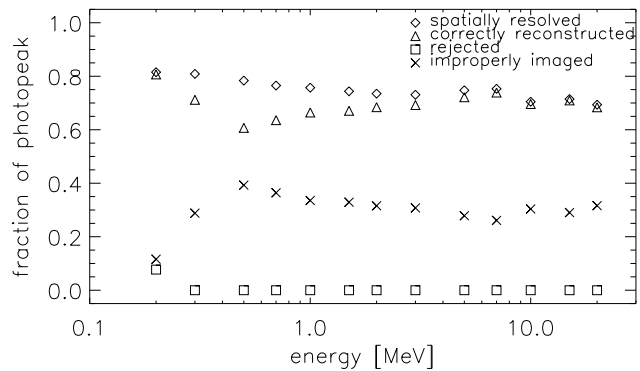}}
  \caption{Photopeak distributions for two-site events. The fraction of events 
   with the first/second interaction sites spatially resolved are presented 
   ($\diamond$), along with the fraction of events which have been properly 
   ordered (hence imaged) using SSD ($\triangle$). The fraction of events rejected 
   due to no physically possible orderings ($\Box$), as well as the fraction 
   incorrectly imaged ($\times$) are also shown.}
  \label{Figure 6}
\end{figure}

\section{Full reconstruction with background rejection}

Given these two methods of determining the photon interaction order in GCTs, CKD 
and SSD, a full event reconstruction technique incorporating other background
rejection techniques can be developed. The high spectral and spatial resolution of a 
GCT make several powerful background rejection techniques possible. The 
predominance of multiple scattering events, while initially a complication,
dramatically helps in the overall background rejection.

The dominant sources of background in GCTs are expected to be diffuse 
cosmic $\gamma$-ray emission, induced satellite $\gamma$-ray emission, and 
induced $\beta^{+}$, $\beta^{-}$ radioactivities in the GeDs themselves
(\cite{jean96, graha97, gehre85, dean91, naya96}).
Source/background photons which scatter out of the instrument before depositing all of 
their energy, and hence are improperly imaged, must also be included in these 
calculations.

Restrictions on the acceptable events can have a dramatic effect
on sensitivity of a GCT. Specifically, several factors can
affect the angular resolution of the instrument as well as the
background rates --
such as the inclusion of backscatter events, limits on the accepted scatter 
angles, and the minimum acceptable lever arm -- and must be included in any 
discussion of full reconstruction and background rejection.

\subsection{Shield veto}

Placing an active shield below the bottom GCT detector plane could be 
useful for rejecting background photons from below the instrument (induced 
satellite $\gamma$-ray emission), as well as helping to reject Compton continuum 
and $\beta-$decay events in the instrument. However, many of these events can be 
distinguished and rejected using CKD and other tests/restrictions outlined below; therefore, 
the usefulness of including a shield in the GCT design must be studied in detail 
for a given telescope configuration.

\subsection{Restrictions on number of interaction sites: pair production/$\beta^{+}$ decays}

While it is obvious that any single-site interactions should be rejected in a 
Compton telescope, events with $\sim$8 or more interaction sites should also be 
rejected since these are very likely due to pair production events, as is evident 
from Figure~3, or similarly $\beta^{+}$ decays. (See Section~6.8)

\subsection{CKD $\chi^{2}_{min}$ test: Compton continuum, unresolved interactions, etc.}

Using the tests outlined in Sections~4 and 5, the most likely ordering of 
the interaction sites can be determined, and for 3+ site events many of the 
unresolved and Compton continuum events, as well as pair production and
$\beta-$decays, rejected.

Shown in Figure~7 is the peak-to-Compton ratio for 3+ site events,
here defined as the ratio of the properly imaged 
photopeak events to the corresponding integrated Compton continuum 
(photons which scatter out of the instrument before depositing all of their 
energy). This standard measure for $\gamma$-ray spectroscopy instruments has 
an altered meaning here, since the Compton continuum events will be 
incorrectly imaged, and thus will appear as off-source background.
The peak-to-Compton ratio is shown both before and after
rejection of the continuum events with the CKD 
statistic. CKD 
rejection of the Compton continuum events increases the
peak-to-Compton ratio by factors of 3-6, an
important improvement for low 
background instruments. By rejection of events appearing to originate from 
below the instrument (Section~6.4), as well as backscattered interactions 
(Section~6.5), this ratio can be increased by further factors of 2-4.

Also shown in Figure~7 is the photopeak-to-Compton ratio for 
two-site events using SSD, which is significantly lower than the corresponding ratio for 
3+ site events. In fact, this ratio drops significantly below unity, which means 
that more background than signal is being created in the instrument for
two-site events. This result questions whether two-site events should be included in 
actual observational analysis given the accompanying increase in unrejectable 
background. This conclusion is further supported by the fact that the majority 
of two-site events are backscatters (Section~6.5), which will significantly 
degrade the angular resolution. Even though inclusion of two-site events is 
unlikely to improve the overall sensitivity for GCTs, detailed background 
analysis for specific instrument configurations is required to determine the 
overall effects.

\begin{figure}
  \resizebox{\hsize}{!}{\includegraphics{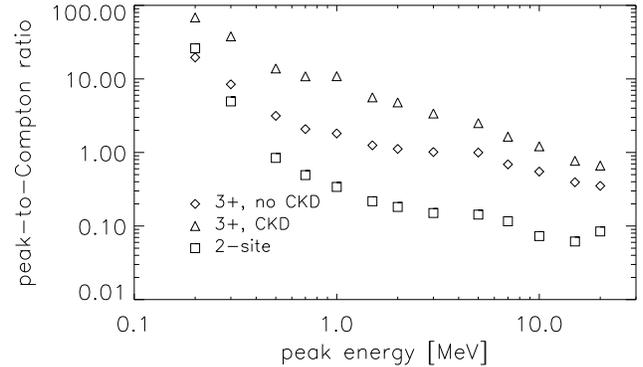}}
  \caption{Peak-to-Compton ratios. Shown are the ratios of correctly
   ordered (imaged) photopeak events to integrated Compton continuum
   events for: 3+ sites before CKD rejection ($\diamond$), 3+ sites
   after CKD rejection ($\triangle$), and two-site events ($\Box$).}
  \label{Figure 7}
\end{figure}

\subsection{Effective TOF}

Once the most likely order of interactions and the initial scatter angle are 
determined, it is possible to determine whether the incident photon scattered 
upwards or downwards in the instrument, as well as whether the initial scatter 
was forward or backwards. Thus, events which appear to be photons originating 
from below the instrument can be rejected, which include the 
induced satellite $\gamma$-ray emission, many Compton continuum events which 
were not rejected by CKD, photons which scatter in the passive satellite
material before interacting in the detectors, and many of
the pair production and $\beta-$decay events. The simulation results for 
the configuration in Appendix~A show that $\sim 95\%$ of photons originating from 
below the instrument are rejected.

\subsection{Backscatters}

Once the most likely ordering of interaction sites has been determined, 
this information can also be used to accept/reject backscattered source photons. 
The fraction of photopeak 
events which backscatter during the initial interaction is not strongly energy 
dependent, $\sim 60-70\%$ for two site events, and $\sim 30\%$ for 3+ site events.
These events can significantly increase the effective area at
lower energies, where two-site events are most common, at 
the expense of degrading the angular resolution due to larger uncertainties in
$\delta \phi_{1,E}$ for backscatters events (Equation 4). It is 
unlikely that the overall sensitivity will improve by including backscattered 
events given the increased background rates and degraded angular resolution; 
however, the effects on sensitivity will depend on the exact instrument
configuration and observational goals.

\subsection{`Standard $\phi$ restriction'}

Restrictions can also be set on the scattering angles accepted for
forward-scattering photons entering the front of the instrument. These limits can be used 
to restrict the instrument FOV to improve imaging capabilities and background, 
such as the `standard $\phi$ restriction' (\cite{schon82}). These restricitions will
have to be reanalyzed in detail for specific GCT configurations.

\subsection{Nonlocalized $\beta^{-}$ decays}

In a nonlocalized $\beta^{-}$ decay, the daughter nuclide is produced
in an excited state which quickly
decays on timescales relative to the detector collection time, emitting a photon 
with energy characteristic to the daughter nuclide.
Therefore, the event consists of the intial $\beta^{-}$ decay site, plus
the interaction sites of the emitted photon.
Such an event can 
be rejected if the characteristic photon energy can be detected in any
combination of the interaction site energies. The seven dominant $\beta^{-}$
isotopes and characteristic photon energies for natural Ge are given in Table~1
(\cite{naya96, gehre85}).
In general, if the coincident $\gamma$-ray is fully deposited, then 
the event will have $\chi^{2}_{min} \sim 1$, with the $\beta^{-}$ electron interaction
ordered as the
initial `scatter' site. In these cases, $W_{1}$ will have the characteristic photon energy 
specific to that decay.
The rejection of all events with $W_{1}$ equal to one of the characteristic energies in
Table~1 can dramatically decrease the $\beta^{-}$ decay background, with only a
small effect, typically $\leq 3\%$ drop in photopeak efficiencies for true photon events.
More  $\beta^{-}$ decays can be rejected if every possible combination of
interaction sites is tested for the decay photon energies, at the expense, however, of
more rejection of photopeak events, typically $\sim 15-20\%$.
If the coincident $\gamma$-ray is only partially deposited, the event will likely be 
rejected by the limits set on $\chi^{2}_{min}$.

\begin{table}
  \caption[]{Characteristic photon energies for the strongest nonlocalized 
             $\beta^{-}$ decays in natural Ge.}
  \[
    \begin{array}{p{0.5\linewidth}l}
     \hline
     \noalign{\smallskip}
     daughter & photon~energy \\
     \noalign{\smallskip}
     \hline
     \noalign{\smallskip}
     $^{75}Ge$ & 0.265~MeV \\
     $^{73}Ga$ & 0.297~MeV \\
     $^{72}Ga$ & 0.834~MeV \\
     $^{71}Zn$ & 0.512~MeV \\
     $^{76}As$ & 0.559~MeV \\
     $^{28}Al$ & 1.779~MeV \\
     $^{77}Ge$ & 0.216~MeV \\
     \noalign{\smallskip}
     \hline
    \end{array}
  \]
\end{table}

$\beta^{-}$ decay background events were simulated for the instrument discussed 
in Appendix~A. The results of these background calculations will be 
presented in a separate paper, but here we make preliminary use of these
simulations to demonstrate the background rejection capabilities. After initial
rejection of two site ($34.4\%$) and 8+ site ($2.9\%$) events,
$62.7\%$ of the nonlocalized
$\beta^{-}$ events remain. Applying the CKD test, requiring a $5\%$ probability of $\chi^{2}_{min}$, 
brings the remaining number of decays down to 17.9\%. After screening the 
interactions for characteristic $\beta^{-}$ decay energies, this number is reduced to 
$9.3\% (15.1\%)$. Finally, after rejecting events which appear to originate from below the 
instrument, or which appear to be backscatter events, the final number of
unrejected $\beta^{-}$ decays comes to $4.2\% (6.8\%)$, a factor of 20 (15) reduction in this background 
component. (First numbers give results when all combinations of interaction sites are
searched for $\beta^{-}$ decay energies, while the numbers in parenthesis are results when
only $W_{1}$ is tested. Typical errors $\sim 0.2\%$.)

\subsection{Positron signatures}

A further test for rejecting the pair production/$\beta^{+}$ background events that 
survive the other tests/restrictions outlined above is 
to search for positron annihilation signatures in the interaction energies. By 
analyzing all combinations of the interaction sites to see if the energies sum to 
$m_{e}c^{2} = 0.511~MeV$, events with a positron annihilation
signature can be rejected. This test typically reduces the non-pair production
photopeak events by $\leq 2\%$.

$\beta^{+}$ background events were simulated for the instrument discussed in 
Appendix A. After initial rejection of two site events ($23.0\%$) and 8+ site 
events ($3.0\%$), $74.0\%$ of the events remain. After the CKD test, $16.7\%$ remain. 
After screening the interactions for 0.511 MeV positron annihilation signatures, 
the number is reduced to $5.2\%$. Finally, after rejecting events which appear to 
originate from below the instrument, or which appear to be backscatter events, 
the final number of unrejected $\beta^{+}$ events is $1.9\%$, a factor of 50 reduction 
in this background component. Similar reductions occur when these tests are 
applied to pair production events. (Typical errors $\sim 0.1\%$.)

\subsection{Minimum lever arm}

In general, a minimum acceptable distance between the first and second 
interaction sites -- the lever arm -- must be set. Figure~8 shows the fraction of 
0.5 and 2.0~MeV photopeak events with lever arms above a given level, for the 
instrument configuration in Appendix~A. Similar to the case of backscattered 
events, a smaller minimum lever arm means a higher effective area at the 
expense of poorer angular resolution. The exact lever arm chosen will depend 
on the instrument configuration and observational goals.

\begin{figure}
  \resizebox{\hsize}{!}{\includegraphics{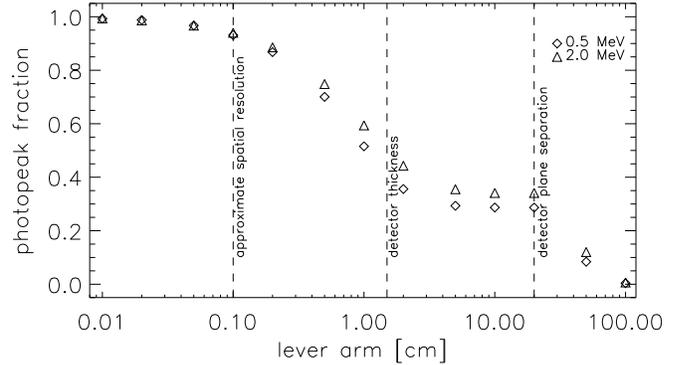}}
  \caption{The fraction of 0.5 and 2.0~MeV photopeak events with lever arms 
  (separations between the first and second interaction sites) greater than the 
  specified values for the model in Appendix~A. Shown for comparison are 
  several characteristic distances of this model.}
  \label{Figure 8}
\end{figure}

\begin{table*}
  \caption[]{Percentage of events remaining after subsequent application of rejection
             techniques.}
  \[
    \begin{array}{p{0.2\linewidth}|*{6}{lccccc}}
     \hline
     \noalign{\smallskip}
     rejection & 0.5 MeV & 2.0 MeV & $$\beta^{-}$$ decays &
     $$\beta^{+}$$ decays & spacecraft \\
     technique & photopeak & photopeak &   &   &  \\
     \noalign{\smallskip}
     \hline
     \noalign{\smallskip}
     2 site events & 65.0\% & 82.0\% & 65.6\% & 77.0\% & 61.6\% \\
     8+ site events & 64.9\% & 80.9\% & 62.7\% & 74.0\% & 60.8\% \\
     CKD & 51.2\% & 61.4\% & 17.9\% & 16.7\% & 35.2\% \\
     $\beta$ signatures & 48.4\% & 58.0\% & 15.1\% & 5.2\% & 32.2\% \\
     backscatter/TOF & 35.5\% & 38.5\% & 6.8\% & 1.9\% & 12.2\% \\
     min lever arm (10 cm) & 15.0\% & 16.7\% & 2.0\% & 1.2\% & 2.6\% \\
     \noalign{\smallskip}
     \hline
    \end{array}
  \]
\end{table*}

\section{Conclusions}

Event reconstruction in future high resolution Compton
telescopes will present a number of complications compared
to historical configurations.
The initial complication of multiple-scattering of photons
in GCTs, however, turns out to be an advantage:
the application of CKD to 3+ site events, combined 
with the high spectral and spatial resolution of GeDs, allows extremely
efficient background suppression, crucial for Compton telescope
performance. This paper has outlined
a set of tests and restrictions, accounting for realistic instrument/detector
performance, to reconstruct photopeak events in GCTs while rejecting a large
fraction of the background events.
Table 2 presents the fraction of events, photon and background, that
remain after each rejection technique is subsequently applied.
(The numbers in Table 2 assume only $W_{1}$ is tested for $\beta^{-}$ decay energies.)
Development of these event reconstruction techniques allows realistic
evaluation of the performace and sensitivity of GCT designs. Our next goal is
to simulate the efficiency, resolution, background and sensitivity of several Compton
telescope configurations, utilizing the event reconstruction techniques 
developed here to realistically determine the performace of these
instruments. CKD rejection has been shown to be the
most efficient background rejection technique; however, the addition of
effective TOF, backscatter, nonlocalized $\beta^{-}$ decay, and positron
signature tests dramatically improve background rejection capabilities. We
anticipate that use of these techniques will achieve
overall sensitivity improvements in GCTs by factors of $\sim 5-10$.

\begin{acknowledgements}

\end{acknowledgements}
S. E. Boggs would like to thank the Millikan Postdoctoral Fellowship Program,
CIT Deparment of Physics, for support.

\appendix
\section{Example GCT configuration}

While it is not the intention of this paper to fully characterize the 
performance of a specific GCT, it is useful to have a telescope model for 
which the results of the event reconstruction can be presented. The telescope 
configuration modeled in this study is presented in Figure~1. The instrument 
consists of five planar arrays of 15~mm thick germanium, each of area 
$100~cm \times 100~cm$. In reality each array would consist of separate smaller 
detectors $(\sim 5~cm \times 5~cm)$ tiled to form the entire plane; however, the 
simulation performed here modeled each plane as a solid detector for 
simplicity. The five planar arrays are spaced 20~cm apart.

This configuration differs from historical Compton telescope 
configurations which generally consist of two detector planes separated by 
100-150~cm. This separation distance is determined by the spatial resolution in 
z and the desired angular resolution. As will be discussed in a second paper, the 
configuration modeled here significantly improves the effective area of the 
telescope by letting each plane act as converter, and permitting a 
much wider range of scatter angles to produce good events. Allowing
large-angle scatters also significantly increases the instrument FOV, and limits the 
effects of point spread function smearing for sources at large off-axis angles. 
The potential drawbacks of this configuration are increased background and 
degraded angular resolution.

The instrument was simulated using CERN's GEANT Monte Carlo code.
The Monte Carlo simulation produces a file of interaction 
locations and energy depositions for each photon/$\beta-$decay event. Before performing 
event reconstruction on the interactions, the simulated events are modified to 
reflect realistic measurement uncertainties of an instrument: for each interaction, a random 
Gaussian-distributed uncertainty is added to the energy and position of each 
interaction. All interaction 
locations which lie within twice the instrumental spatial resolution of each 
other are combined into a single interaction site, to accurately reflect the 
resolving power of the detectors. Finally, interaction sites with energy deposits 
below the assumed detector threshold of 10~keV are ignored.

\begin{figure}
  \resizebox{\hsize}{!}{\includegraphics{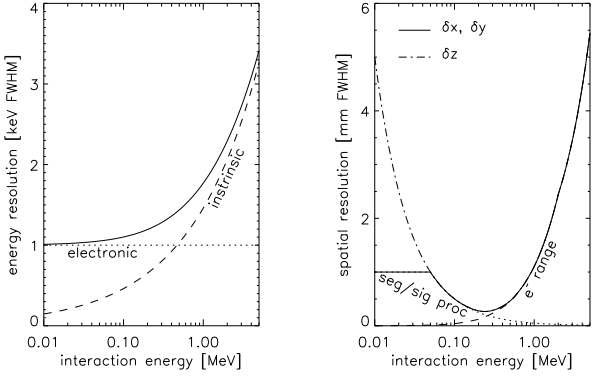}}
  \caption{Assumed resolutions for the GCT model. (Left) The total energy 
  resolution (solid line) is determined by the electronic noise (dotted line) and 
  the intrinsic resolution (dashed line) added in quadrature. (Right) The total x-, 
  y- (solid line) and z- (dot-dashed line) spatial resolution are determined by the 
  segmentation/signal processing limits (dotted line), and the recoil electron 
  range (dashed line) added in quadrature. (See text for details.)}
  \label{Figure A1}
\end{figure}

Two components are assumed to add in quadrature to determine the 
energy resolution: (i) a constant electronic noise, $W_{e} = 1.0~keV~FWHM$, and 
(ii) the intrinsic resolution $W_{i}$ determined by the germanium Fano factor,
$F = 0.13$, and average free electron-hole pair energy, $\varepsilon = 2.98~eV$, giving 
$W_{i} = 2.35 \sqrt{F \varepsilon E}~FWHM$. This corresponds to a resolution  $\sim 1.8~keV~FWHM$
at 1~MeV, which is optimistic but not unrealistic. It is assumed that 
charge trapping and ballistic deficit do not significantly alter this energy 
resolution. The two components as well as the total energy resolution are 
shown in Figure~A1.

It is assumed that two components add in quadrature to determine the $1-D$
spatial resolutions, $\delta x, \delta y, \delta z$, of the detectors:
(i) the range of the recoil 
electrons in the detector, and (ii) the positioning limits of the detector due to 
physical segmentation and/or signal analysis. Calculated electron ranges in
germanium for different energies (\cite{mukoy76}) are used
here as the 1-D FWHM positional uncertainties,
$\delta x_{e}, \delta y_{e}, \delta z_{e}$. Methods to 
determine the event position by physically segmenting the GeD contacts into 
cross strips or pixels (\cite{luke94, kroeg96}), as well as using advanced signal 
processing to interpolate to even better positions (\cite{boggs98, luke94}),
are currently active fields of research -- so this component of the 
spatial resolution remains speculative for now. Here it is assumed that signal 
processing will allow positional resolutions of $\sim 0.5~mm~FWHM$ at 100~keV, 
and that the discrimination capabilities go as the signal-to-noise ratio of the 
induced detector signal to electronic noise, i.e. as the inverse power of the 
interaction energy. It is also assumed that there is $\sim 1~mm$ physical 
segmentation of the detector contacts in x, y, so that this component never 
exceeds this value. The z uncertainty, however, is not constrained by
any such segmentation 
at the lowest energies. Therefore, the signal processing uncertainty is given by 
$\delta x_{s}, \delta y_{s}, \delta z_{s} \sim 0.50 (E/100 keV)^{-1}~mm~FWHM$,
maximizing at 1~mm in x, y below 50 keV,
and approaching, but never maximizing at 15~mm in z at low energies.
The two components as well as the total spatial resolution are 
shown in Figure~A1.


\begin{thebibliography}{}


   \bibitem[Aprile et al. 1993]{april93} Aprile E., Bolotnikov A., Chen D.,
      Mukherjee R., 1993, Nucl. Inst. \& Methods A 327, 216

   \bibitem[Boggs 1998]{boggs98} Boggs S. E., 1998,
      Ph. D. Dissertation, University of California, Berkeley

   \bibitem[Dean et al. 1991]{dean91} Dean A. J., Lei F., Knight P. J., 1991,
      Space Sci. Rev. 57, 109

   \bibitem[Gehrels 1985]{gehre85} Gehrels N., 1985, Nucl. Inst. \& Methods A 239, 324

   \bibitem[Graham et al. 1997]{graha97} Graham B. L., Phlips B. F., Kurfess J. D.,
      Kroeger R. A., 1997, IEEE Nucl. Sci. Symp. N23-3

   \bibitem[Herzo et al. 1975]{herzo75} Herzo D., et al. 1975,
      Nucl. Inst. \& Meth. 123, 583
 
   \bibitem[Jean et al. 1996]{jean96} Jean P., Naya J. E., Olive J. F.,
      von Ballmoos P., 1996, A\&ASS 120, 673

   \bibitem[Johnson et al. 1996]{johns96} Johnson W. N., et al., 1996,
      SPIE

   \bibitem[Kroeger et al. 1996]{kroeg96} Kroeger R. A., et al., 1996, IEEE

   \bibitem[Lichti et al. 1996]{licht96} Lichti G. G., et al., 1996,
      SPIE 2806, 217

   \bibitem[Luke et al. 1994]{luke94} Luke P. N., et al, 1994, IEEE

   \bibitem[Lockwood et al. 1979]{lockw79} Lockwood J. A., Hsieh L., Friling L.,
      Chen C. \& Swartz D., 1979, JGR 84, 1402

   \bibitem[Mukoyama 1976]{mukoy76} Mukoyama, T., 1976, Nucl. Inst. \& Meth. 134, 125

   \bibitem[Naya et al. 1996]{naya96} Naya J. E., et al., 1996, Nucl. Inst. \& Meth. A 368, 832

   \bibitem[Sch\"onfelder et al. 1973]{schon73} Sch\"onfelder V., Hirner A.,
      \& Schneider K., 1973, Nucl. Inst. \& Meth. 107, 385

   \bibitem[Sch\"onfelder et al. 1982]{schon82} Sch\"onfelder V., Graser U.,
      \& Diehl R., 1982, A\&A 110, 138

   \bibitem[Sch\"onfelder et al. 1993]{schon93} Sch\"onfelder V., et al. 1993,
      ApS 86, 657

   \bibitem[Vedrenne et al. 1998]{vedre98} Vedrenne G., Sch\"onfelder V.,
      et al., 1998, Proc. 3rd INTEGRAL Workshop, Taormina

   \bibitem[von Ballmoos et al. 1989]{vonba89} von Ballmoos P., Diehl R., \& Sch\"onfelder V.,
      1989, A\&A 221, 396

\end{thebibliography}
\end{document}